\documentclass[twocolumn]{revtex4}
\usepackage{epsfig}
\usepackage{graphicx}
\usepackage{amssymb}
\usepackage{amsmath}
\usepackage{amsthm}
\usepackage{amsfonts}
\usepackage{mathrsfs}
\usepackage[dvips]{color}
\usepackage{multirow}

\newcommand{\fn}{{\,\mathfrak{n}\,}}

\newcommand{\bM}{\mathbf{M}}
\newcommand{\bN}{\mathbf{N}}

\newcommand{\cR}{\mathcal{R}}

\newcommand{\be}{\begin{equation}}
\newcommand{\ee}{\end{equation}}
\newcommand{\bea}{\begin{eqnarray}}
\newcommand{\eea}{\end{eqnarray}}

\newcommand{\ed}{\end{document}}

\newcommand{\bi}{\begin{itemize}}
\newcommand{\ei}{\end{itemize}}

\newcommand{\bce}{\begin{center}}
\newcommand{\ece}{\end{center}}

\newcommand{\sR}{\mathscr{R}}

\newcommand{\sT}{\mathscr{T}}

\newcommand{\fno}{\fn_{\!0}}
\newcommand{\fnone}{\fn_{\!1}}

\begin{document}

\title{Active Invisibility Cloaks in One Dimension}

\author{Ali~Mostafazadeh}
\address{Departments of Physics and Mathematics, Ko\c{c} University, Sar{\i}yer 34450, Istanbul, Turkey\\
amostafazadeh@ku.edu.tr}

\begin{abstract}

We outline a general method of constructing finite-range cloaking potentials which render a given finite-range real or complex potential $v(x)$ unidirectionally reflectionless or invisible at a wavenumber $k_0$ of our choice. We give explicit analytic expressions for three classes of cloaking potentials which achieve this goal while preserving some or all of the other scattering properties of $v(x)$. The cloaking potentials we construct are the sum of up to three constituent unidirectionally invisible potentials. We discuss their utility in making $v(x)$ bidirectionally invisible at $k_0$, and demonstrate the application of our method to obtain anti-reflection and invisibility cloaks for a Bragg reflector.\\

\noindent {Pacs numbers: 42.25.Bs, 03.65.Nk, 02.30.Zz}\\


\end{abstract}

\maketitle

\section{Introduction}

The possibility of making a scatterer invisible to a distant observer by placing a cloaking device between the two has intrigued both scientists and the laymen for a very long time. The discovery of the methods of conformal mapping \cite{Leonhardt} and transformation optics \cite{pendry} combined with the mind boggling possibilities offered by metamaterials have recently led to a rapid progress towards the realization of such invisibility cloaks in two and three-dimensions \cite{InvC}. The one-dimensional analogue of this problem is typically considered in the context of developing antireflection coatings \cite{yeh,southwell}. It involves constructing a finite range potential $v_c(x)$ whose addition to a given finite-range target potential yields a unidircetionally reflectionless or invisible total potential. The purpose of the present article is to give an exact and analytic solution of this problem in such a way that the presence of $v_c(x)$ does not alter some or any of the scattering properties of the system in the direction along which it remains visible.

The construction of one-dimensional antireflection cloaks is intimately related to the old problem of characterizing reflectionless potentials. In Ref.~\cite{kay-moses}, Kay and Moses use the powerful tools of inverse scattering theory \cite{CS} to provide a systematic method of constructing an infinite class of potentials which are reflectionless at all wavenumbers. The best-known member of this class is the reflectionless P\"oschl-Teller potential \cite{flugge-v1}. The authors of Refs.~\cite{gupta-2007,thekkekara} employ the results of Kay and Moses to devise a method of constructing realistic broadband antireflection coatings. Here a serious practical problem is that the reflectionless potentials of Kay and Moses have an infinite-range. Therefore, they cannot be realized using material (refractive index profiles) confined to a closed region in space.

Complex absorbing potentials \cite{muga,moiseyev} offer another means of realizing an antireflection cloak in one-dimension. By definition, an absorbing potential has zero (left or right) reflection and transmission coefficients  \cite{muga}. This implies that it may function as an antireflection cloak from the left (respectively right), but its presence will drastically alter the scattering properties of the system from the right (respectively left). In order to provide a more detailed assessment of the utility of absorbing potentials in realizing invisibility cloaks and offer a precise formulation of the problem we consider in this article, we briefly survey some of the basic notions of one-dimensional scattering theory.

By definition, the reflection and transmission amplitudes, $R^{l/r}(k)$ and $T(k)$, of a scattering potential $v(x)$ are complex numerical quantities which determine the asymptotic form of the left- and right-incident scattering solutions $\psi_{l/r}(x)$ of the schr\"odinger equation,
    \be
    -\psi''(x)+v(x)\psi(x)=k^2\psi(x),
    \label{sch-eq}
    \ee
according to
    \bea
    \psi_l(x)&=&\left\{\begin{array}{ccc}
    e^{ikx}+R^l(k)\,e^{-ikx}&{\rm for}& x\to-\infty,\\
    T(k)\,e^{ikx}&{\rm for}& x\to+\infty,\end{array}\right.
    \label{psi-L}\\
    \psi_r(x)&=&\left\{\begin{array}{ccc}
    T(k)\,e^{-ikx}&{\rm for}& x\to-\infty,\\
    e^{-ikx}+R^r(k)\,e^{ikx}&{\rm for}& x\to+\infty.\end{array}\right.
    \label{psi-R}
    \eea
Equations~(\ref{psi-L}) and (\ref{psi-R}) also apply for
the Helmholtz equation, $\psi''(x)+k^2\fn(x)^2\psi(x)=0$, which coincides with (\ref{sch-eq}) if we relate the refractive index $\fn(x)$ to the potential $v(x)$ via $\fn(x)=\sqrt{1-v(x)/k^2}$.

For simplicity of presentation, we choose an arbitrary but fixed value of the wavenumber $k$, which we label by $k_0$, and use the symbols $T$ and $R^{l/r}$ to denote $T(k_0)$ and $R^{l/r}(k_0)$, respectively. Note also that we label both the left and right transmission amplitudes by $T$, because they coincide \cite{jpa-2009}.

A scattering potential is said to be unidirectionally invisible from the left (respectively right) if $R^l=0\neq R^r$ (respectively $R^r=0\neq R^l$) and $T=1$, \cite{lin}. We use the term `left-invisible' (respectively `right-invisible') to refer to such a potential. If a potential is both left- and right-invisible, we call it bidirectionally invisible. Relaxing the condition $T=1$, we similarly define the notions of unidirectional, left-, right-, and bidirectional reflectionlessness.

The study of unidirectionally invisible potentials \cite{invisible-1,lin,invisible-2} has recently attracted a great deal of attention, because they offer an interesting method of modeling certain one-way optical devices \cite{lin}. An equally interesting motivation for exploring these potentials
is their unique role in a recently proposed inverse scattering prescription which allows for the construction of finite-range potentials supporting scattering properties of one's choice at any prescribed wavenumber \cite{pra-2014b}. The following theorem provides a precise statement of this result.\\[6pt]
    \noindent {\em Theorem~1:} Let $k_0$ be a positive real number, and $\sR^l$, $\sR^r$, and $\sT$ be arbitrary complex numbers such that $\sT\neq 0$. Then there is a finite-range potential $v (x)$ with the following properties.
        \begin{enumerate}
        \item The reflection and transmission amplitudes of $v(x)$ at $k=k_0$ are respectively given by $R^{l/r}=\sR^{l,r}$ and $T=\sT$;
        \item If $\sR^l=\sR^r=0$ and $\sT\neq 1$, $v(x)$ is the sum of four unidirectionally invisible finite-range potentials with mutually disjoint support \cite{f1}.
        \item If $|\sR^l|+|\sR^r|\neq 0$, $v(x)$ is the sum of at most three unidirectionally invisible finite-range potentials with mutually disjoint support.
        \item The support of $v(x)$ can be chosen to be on the left or the right of any point on the real line.
        \end{enumerate}

The main technical tool used in the proof of this theorem \cite{pra-2014b} is the transfer matrix of the one-dimensional potential scattering \cite{sanchez,prl-2009}. Every solution of the Schr\"odinger equation (\ref{sch-eq}), for a scattering potential $v(x)$, has the asymptotic form:
    \[\psi(x)=A_\pm e^{ikx} +B_\pm e^{-ikx}~~~~{\rm as}~~~~x\to\pm\infty,\]
where $A_\pm$ and $B_\pm$ are complex coefficients. The transfer matrix of $v(x)$ is, by definition, the $2\times 2$ matrix $\bM$ satisfying
    \[\left[\begin{array}{c}
    A_+\\ B_+\end{array}\right]=\bM \left[\begin{array}{c}
    A_-\\ B_-\end{array}\right].\]
We can express it in terms of the reflection and transmission amplitudes of $v(x)$ according to \cite{prl-2009}
    \be
    \bM=\left[\begin{array}{cc}
    T- R^lR^r/T & R^r/T\\
    - R^l/T & 1/T\end{array}\right].
    \label{M=}
    \ee

Transfer matrices are extremely useful because of their composition property: Suppose that $v_\pm (x)$ are scattering potentials with support $I_\pm$ and transfer matrix $\bM_\pm$. If $I_-$ lies to the left of $I_+$ (which we denote by $I_-\prec I_+$), the transfer matrix of $v_-(x)+v_+(x)$ is given by
	\be
	\bM=\bM_+\bM_-.
	\label{group}
	\ee
We can use this relation and (\ref{M=}) to express the reflection and transmission amplitudes, $R^{r/l}$ and $T$, of $v_-(x)+v_+(x)$ in terms of those of $v_\pm(x)$. Denoting the latter by $R_\pm^{l/r}$ and $T_\pm$ and using (\ref{M=}) and (\ref{group}), we have
    \bea
    R^l&=&\frac{(T_-^2-R^l_-R^r_-)R^l_++R^l_-}{1-R_-^rR_+^l},
    \label{zqL}\\
    R^r&=&\frac{(T_+^2-R^l_+R^r_+)R^r_-+R^r_+}{1-R_-^rR_+^l},
    \label{zqR}\\
    T&=&\frac{T_-T_+}{1-R_-^rR_+^l}.
    \label{zqT}
    \eea

Suppose that $v(x)$ is a given finite-range potential. If $v_c(x)$ is a finite-range potential whose support does not intersect that of $v(x)$ and the addition of  $v_c(x)$ to $v(x)$ yields a unidirectionally or bidirectionally invisible total potential, $v(x)+v_c(x)$, we say that $v_c(x)$ is an invisibility cloak for $v(x)$. Similarly, we call $v_c(x)$ an antireflection cloak, if $v(x)+v_c(x)$ is reflectionless.

For example, consider adding an absorbing potential $v_-(x)$, with $R^l_-=T_-=0$, to a target potential $v_+(x)$ such that $I_-\prec I_+$. Then according to (\ref{zqL}) and (\ref{zqT}), the left reflection and transmission amplitudes of $v_-(x)+v_+(x)$ vanish. Therefore, $v_-(x)+v_+(x)$ is left-reflectionless, but not left-invisible. This argument shows that the absorbing potentials can never be used to construct genuine invisibility cloaks.

Next, consider an antireflection cloaking potential $v_c$, so that the left (or right) reflection amplitude of $v(x)+v_c(x)$ vanishes. If the addition of $v_c(x)$ to $v(x)$ leaves both the right (respectively left) reflection amplitude and the transmission amplitude of $v(x)$ intact, we say that $v_c(x)$ is an optimal antireflection cloak (or cloaking potential) for $v(x)$. Similarly, we define the notion of an optimal invisibility cloak by demanding that its addition does not alter the right (respectively left) reflection amplitude of $v(x)$. According to this terminology, the absorbing potentials yield non-optimal antireflection cloaks. In this article, we give a general method of constructing optimal antireflection and invisibility cloaks for an arbitrary finite-range target potential.

We close this section by summarizing the construction of a concrete model for unidirectionally invisible potentials \cite{pra-2014b} that we use extensively in the remainder of this article. The importance of such a model is clearly highlighted by Theorem~1.

Given a nonzero complex number $\cR$, let
	\begin{align}
	&v^r_\cR(x):=\left\{\begin{array}{cc}
	k^2 f_{\alpha}(x-a_m)&{\rm for}~x\in[a_m,L_n+a_m],\\[6pt]
	0&{\rm for}~x\notin[a_m,L_n+a_m],\end{array}\right.
	\label{model-r}\\[6pt]
	& v^l_\cR(x):=v^r_{-\cR^*}(x)^*,
	\label{model-l}
	\end{align}
where
    \bea
	&&f_{\alpha}(x):=\frac{-8\,\alpha (3-2 e^{2ik_0x})}{e^{4ik_0x}+\alpha(1-e^{2ik_0x})^2},
	\label{f=}\\
	&& a_m:=\frac{(4m+1)\pi+2\varphi}{4k_0},
	\label{d=}
	\eea
$m$ is an arbitrary integer, $n$ and $\alpha$ are respectively a positive integer and a positive real number satisfying
	\be
	|\cR|(\alpha+1)^3-8\pi n\alpha=0,
	\label{e01}
	\ee
$L_n:=\pi n/k_0$, and $\varphi$ is the principal argument (phase angle) of $\cR$, i.e., $\varphi=-i\ln(\cR/|\cR|)\in[0,2\pi)$. Then, $v^{l}_{\cR}(x)$ (respectively $v^{r}_{\cR}(x)$) is a left-invisible (respectively right-invisible) potential with the right (respectively left) reflection amplitude $\cR$, \cite{pra-2014b}. Note that because $m$ can take arbitrary integer values, we can choose it so that the support of $v^{r/l}_{\cR}(x)$ lies to the left or right of any point or finite interval.

\section{Optimal Unidirectional Invisibility Cloaks}

Suppose that $v(x)$ is an arbitrary finite-range potential with support $I$ and reflection and transmission amplitudes, $R^{l/r}$ and $T$, so that its transfer matrix $\bM$ is given by (\ref{M=}). Theorem~1 implies the existence of finite-range potentials $u_\pm(x)$ with support
$I_\pm$ and reflection and transmission amplitudes, $R^{l/r}_\pm$ and $T_\pm$, such that $I_-\prec I \prec I_+$ and (for $k=k_0$),
	\begin{align}
	&R_-^l=-\frac{R^l}{T^2}, && R_-^r=0, && T_-=\frac{1}{T},
	\label{eqm}\\
	&R_+^l=0, && R_+^r=-\frac{R^r}{T^2}, && T_+=\frac{1}{T}.
	\end{align}
In view of (\ref{M=}), the transfer matrix $\bM_\pm$ of $u_\pm(x)$ at $k_0$ takes the form
	\begin{align}
	&\bM_-=\left[\begin{array}{cc}
	1/T & 0 \\
	R^l/T & T\end{array}\right],
	&& \bM_+=\left[\begin{array}{cc}
	1/T & -R^r/T \\
	0 & T\end{array}\right].
	\label{M-mp}
	\end{align}
Because, $I_-\prec I \prec I_+$, the composition property of the transfer matrix implies that the transfer matrix of $v(x)+u_-(x)$ and $v(x)+u_+(x)$ are respectively given by
    \begin{align}
	& \bM\bM_-=\left[\begin{array}{cc}
	1 & R^r\\
	0 & 1\end{array}\right],
	&& \bM_+\bM=\left[\begin{array}{cc}
	1& 0 \\
	-R^l & 1\end{array}\right].
    \label{MM-mp}
	\end{align}
This shows that $v(x)+u_-(x)$ is left-invisible and has the same right reflection amplitude as $v(x)$, and $v(x)+u_+(x)$ is right-invisible and has the same left reflection amplitude as $v(x)$.
Therefore, $u_\pm(x)$ are unidirectionally reflectionless potentials that serve as optimal invisibility cloaks for $v(x)$.

Suppose that not both $R^l$ and $R^r$ vanish. Then, according to Theorem~1, we can construct $u_\pm(x)$ using three unidirectionally invisible  potentials, which we denote by $u_{j\pm}(x)$ with $j=1,2,3$.

We identify $u_{j\pm}(x)$ with a finite-range potential having the domain $I_{j\pm}$ and the transfer matrix $\bM_{j\pm}$ such that $I_{1\pm }\prec I_{2\pm }\prec I_{3\pm}$ and
    \bea
    &&\bM_{1-}=\left[\begin{array}{cc}
	1 & \displaystyle\frac{T(T-1)}{R^l}\\[6pt]
	0 & 1\end{array}\right],~~~~
	\bM_{2-}=\left[\begin{array}{cc}
	1& 0 \\
	\displaystyle\frac{R^l}{T} & 1\end{array}\right],
    \label{Mj-1}\\
    &&\bM_{3-}=\left[\begin{array}{cc}
	1 & \displaystyle\frac{1-T}{R^l}\\[6pt]
	0 & 1\end{array}\right],~~~~
    \bM_{1+}=\left[\begin{array}{cc}
	1& 0 \\
	\displaystyle\frac{T-1}{R^r} & 1\end{array}\right],
    \label{Mj-2}\\
    && \bM_{2+}=\left[\begin{array}{cc}
	1 & -\displaystyle\frac{R^r}{T}\\[6pt]
	0 & 1\end{array}\right],~~~~
    \bM_{3+}=\left[\begin{array}{cc}
	1& 0 \\
	\displaystyle\frac{T(1-T)}{R^r} & 1\end{array}\right].~~~~~~
    \label{Mj-3}
	\eea
Then, it is easy to check that $\bM_{\pm 3}\bM_{\pm 2}\bM_{\pm 1}=\bM_\pm$. This shows that we can take
    \be
    u_\pm(x)= u_{1\pm}(x)+u_{2\pm}(x)+u_{3\pm}(x).
    \label{u=uuu}
    \ee
Notice that the construction of the optimal cloaking potential $u_-(x)$ (respectively $u_+(x)$) is desirable only if $R^l\neq 0$ (respectively $R^r\neq 0$). According to (\ref{Mj-1}) -- (\ref{Mj-3}), $\bM_{j-}$ (respectively $\bM_{j+}$) exists precisely for this case.

The above analysis reduces the construction of the optimal invisibility clocking potentials $u_\pm(x)$ to that of the finite-range unidirectionally invisible potentials $u_{j\pm}(x)$. We can employ the model introduced in (\ref{model-r}) and (\ref{model-l}) to give explicit expressions for the latter. The potentials $u_\pm(x)$ that we obtain in this way define optimal invisibility cloaks that eliminate the reflection of the plane waves with wavenumber $k_0$ from one direction, set the transmission amplitude to unity, but do not affect the reflection amplitude of $v(x)$ from the other direction. These invisibility cloaks correspond to planar slabs consisting of three optically active layers with particular locally periodic gain-loss profile and adjustable gaps in between.

\section{Non-Optimal Unidirectional Invisibility Cloaks}

In this section we examine cloaking potentials that render the original potential invisible from one direction but do alter its reflection feature from the other direction.

Consider the finite range potentials $\check u_{\pm \ell}(x)$ with $\ell=1,2$ such that their support $\check I_{\pm\ell}$ and transfer matrix $\check \bM_{\pm\ell}$ satisfy
    \begin{align}
    & \check I_{1-}\prec \check I_{2-}\prec I \prec \check I_{1+}\prec \check I_{2+},
    \label{q100}\\
    &\check\bM_{1-}=\left[\begin{array}{cc}
    1 & 0\\
    \displaystyle \frac{R^l}{T} & 1\end{array}\right],~~~~
    \check\bM_{2-}=\left[\begin{array}{cc}
    1 & \displaystyle \frac{1-T}{R^l}\\[6pt]
    0 & 1\end{array}\right],
    \label{q101}\\
    &\check\bM_{1+}=\left[\begin{array}{cc}
    1 & 0\\
    \displaystyle \frac{T-1}{R^r} & 1\end{array}\right],~~~~
    \check\bM_{2+}=\left[\begin{array}{cc}
    1 & \displaystyle -\frac{R^r}{T}\\[6pt]
    0 & 1\end{array}\right].
    \label{q102}
    \end{align}
Comparing these relations with (\ref{M=}), we see that $\check u_{\pm \ell}(x)$ are unidirectionally invisible potentials with reflection amplitude $\check R^{l/r}_{j\pm}$ given by $\check R^r_{1\pm}=\check R^l_{2\pm}=0$ and
    \begin{align}
    &\check R^l_{1-}=-\frac{R^l}{T}, && \check R^r_{2-}=\frac{1-T}{R^l},
    \label{q103}\\
    &\check R^l_{1+}=\frac{1-T}{R^r}, && \check R^r_{2+}=-\frac{R^r}{T}.
    \label{q104}
    \end{align}

Now, let $\check u_\pm(x):=\check u_{1\pm}(x)+\check u_{2\pm}(x)$. Then, in view of Eqs.~(\ref{q100}) -- (\ref{q102}), the transfer matrix of the potentials $v(x)+\check u_-(x)$ and $v(x)+\check u_+(x)$ are respectively given by
    \bea
    \bM\,\check\bM_{2-} \check\bM_{1-}&=&\left[\begin{array}{cc}
    1~~~~~ & \displaystyle R^r-\frac{T(T-1)}{R^l}\\[6pt]
    0 & 1\end{array}\right],
    \label{q201}\\
    \check\bM_{2+} \check\bM_{1+}\bM &=&\left[\begin{array}{cc}
    1~~~~~ & 0\\
    \displaystyle  -R^l+\frac{T(T-1)}{R^r} & 1\end{array}\right].
    \label{q201}
    \eea
These in turn identify $\check u_\pm(x)$ with a pair of invisibility cloaks which are not necessarily optimal; the addition of $\check u_-(x)$ makes $v(x)$ left-invisible but changes its right reflection amplitude to $R^r-T(T-1)/R^l$, while the addition of $\check u_+(x)$ makes $v(x)$ right-invisible and changes its left reflection amplitude to $R^l-T(T-1)/R^r$.

Again we can use the model described by (\ref{model-r}) and (\ref{model-l}) to give explicit formulas for the constituent unidirectionally invisible potentials $\check u_{\ell\pm}(x)$ and consequently the cloaking potentials $\check u_\pm(x)$. The advantage of the latter over their optimal analogues, namely $u_\pm(x)$, is that they correspond to planar slabs consisting of two optically active layers.

\section{Optimal Unidirectional Antireflection Cloaks and Bidirectional Invisibility}

It is not difficult to see that we can construct (single-layer) cloaking slabs described by unidirectionally invisible potentials $w_\pm(x)$ whose addition to $v(x)$ yields a unidirectionally reflectionless potential. As suggested by (\ref{MM-mp}), we can identify $w_\pm(x)$ with the finite-range potentials whose support $J_\pm$ and transfer matrix $\bN_\pm$ fulfil $J_-\prec I\prec J_+$ and
    \begin{align}
    & \bN_-=\left[\begin{array}{cc}
    1 & 0 \\
    R^l & 1\end{array}\right], &&
    \bN_+=\left[\begin{array}{cc}
    1 & -R^r \\
    0 & 1\end{array}\right],
    \label{q301}
    \end{align}
i.e., $w_-(x)$ is a right-invisible potential with left-reflection amplitude $-R^l$, while $w_+(x)$ is a left-invisible potential with right-reflection amplitude $-R^r$.

Using (\ref{M=}) and (\ref{q301}) to compute the transfer matrix of $v(x)+w_\pm(x)$, we find that the $v(x)+w_-(x)$ (respectively $v(x)+w_+(x)$) is a left (respectively right) reflectionless potential with the same right (respectively left) reflection and transmission amplitudes as $v(x)$. Therefore, $w_\pm(x)$ are optimal reflectionless cloaking potentials. A concrete choice for $w_-(x)$ (respectively $w_+(x)$) is the potential $v^r_{-R^l}(x)$ of Eq.~(\ref{model-r}) (respectively $v^l_{-R^r}(x)$ of Eq.~(\ref{model-l}).)

Next, consider using the cloaking potentials $\hat u_\pm(x)$ together with $w_\mp(x)$. If we adjust the parameters of the latter so that the left (respectively right) reflection amplitude of $w_-(x)$ (respectively $w_+(x)$) coincides with $-R^r+T(T-1)/R^l$ (respectively $-R^l+T(T-1)/R^r$), both the potentials $v(x)+\hat u_\pm(x)+w_\mp(x)$ become bidirectionally invisible, i.e., their transfer matrix coincides with the identity, at $k=k_0$. Notice that the support of $v(x)$ lies between those of $\hat u_\pm(x)$ and $w_\mp(x)$; more specifically, we have $\check I_{1-}\prec \check I_{2-}\prec I \prec J_+$ and $J_-\prec I\prec \check I_{1+}\prec \check I_{2+}$.

\section{Application to a Bragg Reflector}
\label{SecV}

Consider the application of the results of the preceding sections in the construction of invisibility and antireflection cloaks for the following locally periodic potential.
	\be
	v_b(x):=\left\{\begin{array}{cc}
	k^2\big\{1-[\fno + \fnone\sin(K x)]^2\big\}&{\rm for}~x\in[0,L],\\
	0 & {\rm for}~x\notin[0,L],
	\end{array}\right.
	\label{bragg}
	\ee
where $\fno=2.29$, $\fnone=0.01$, $K=2\pi/\Lambda$, $\Lambda=232\,{\rm nm}$, and $L=250\Lambda=58~\mu{\rm m}$. This corresponds to a 1.06~$\mu$m Bragg reflector made out of ZnS \cite{yeh}. Our aim is to construct various cloaking potentials to achieve unidirectional reflectionlessness and invisibility of the cloaked system at the wavelength $\lambda_0=1064~{\rm nm}$. Because of the symmetry of the problem, we confine our attention to the construction of left-invisibility and left-antireflection cloaking potentials.

We can easily use the results of Refs.~\cite{ap-2014,griffiths} to compute the  reflection and transmission amplitudes of the potential (\ref{bragg}) numerically. For $\lambda_0=1064~{\rm nm}$, which corresponds to $k_0=5.90525/\mu{\rm m}$, they are given by
	\begin{align}
	&R_b^l=-0.9102811+0.2133993\,i,
    \label{RL=b}\\
    &R_b^r =-0.4876307+0.7977251\,i
    \label{RR=b}\\
    & T_b =-0.2402088+0.2610598\,i.
	\label{RRT=b}
	\end{align}
In particular, for the reflection and transmission coefficients, $|R_b^{l/r}|^2$ and  $|T_b|^2$, we find
    \be
    |R_b^{l/r} |^2=1-|T_b|^2\approx 87 \%. 
    \label{RT2}
    \ee

\subsection{Antireflection cloaking potential $w_-(x)$}
\label{subsec-4A}

As we mentioned above, a concrete realization of $w_-(x)$ is provided by the potentials of the form (\ref{model-r}) with $\cR=-R_b^l$, i.e.,
    \be
    w_-(x)=v^r_{-R_b^l}(x).
    \label{q311}
    \ee
We only need to make proper choices for the parameters $m$, $n$, and $\alpha$ appearing in the expression for $v^r_{-R_b^l}(x)$.

\begin{figure}
	\begin{center}
	\includegraphics[scale=.6]{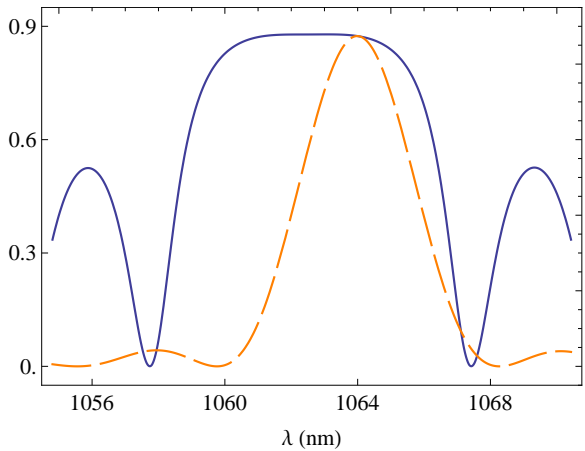}
	\caption{Graphs of the left reflection coefficient $|R^l|^2$ for the Bragg reflector potential $v_b(x)$ (navy, solid curve) and the antireflection cloaking potential $w_-(x)$ (orange, dashed curve) of Sec.~\ref{subsec-4A}.}
	\label{fig1}
	\end{center}
	\end{figure}

Figure~\ref{fig1} shows plots of the left reflection coefficient for the Bragg reflector potential (\ref{bragg}) and the cloaking potential (\ref{q311}) for $m=-19048$, $n=250$, and $\alpha=1.488700\times 10^{-4}$. The latter corresponds to a 133~$\mu$m-thick slab placed at a distance of about 1~cm to the left of the Bragg reflector. Note that by taking smaller and larger values of $m$ we can adjust the distance between the Bragg reflector and the clocking slab.  A remarkable property of the cloaking potential (\ref{q311}) is that it displays broadband right-invisibility; in the spectral range depicted in Fig.~\ref{fig1}, $|R^r|$ and $|T-1|$ are respectively bounded by  $9\times 10^{-7}$ and $6\times 10^{-6}$.

\begin{figure}
	\begin{center}
	\includegraphics[scale=.6]{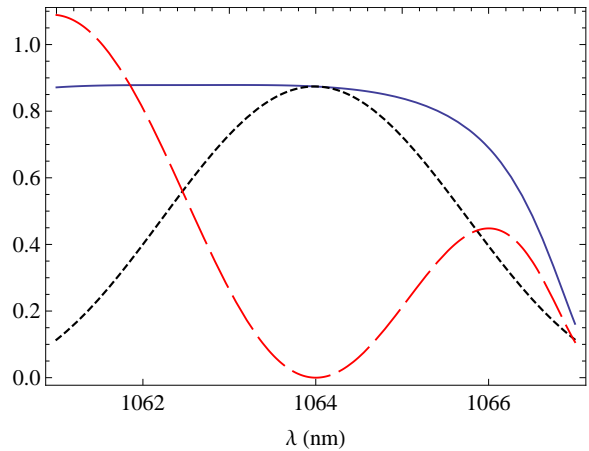}
	\caption{(Color online) Graphs of the left reflection coefficient $|R^l|^2$ for the Bragg reflector potential $v_b(x)$ (navy solid curve), the antireflection cloaking potential $w_-(x)$ (black dotted curve), and the cloaked potential $v_b(x)+w_-(x)$ (red dashed curve) of Sec.~\ref{subsec-4A}.}
	\label{fig2}
	\end{center}
	\end{figure}

Figure~\ref{fig2} shows the graphs of the left reflection coefficient for the Bragg reflector potential $v_b(x)$, the clocking potential $w_-(x)$, and $v_b(x)+w_-(x)$ for the above-given choice of the parameters $n,m$, and $\alpha$. The cloaking effect happens to be restricted to a narrow spectral band of less than 1~nm in width. Notice however that this is an extreme example where we attempt to make a mirror reflectionless.
For the wavelengths outside the range $[1054\,{\rm nm},1074\,{\rm nm}]$, the cloaking potential is essentially bidirectionally invisible and the reflection and transmission coefficients of $v_b(x)+w_-(x)$ coincide with those of $v_b(x)$. Within this range, their left reflection coefficient differ appreciably, but their right reflection coefficient and the transmission coefficient agree to an extremely high degree of accuracy (Their difference is less than $10^{-5}$.)

\subsection{Non-optimal invisibility cloaking potential $\check u_-(x)$}
\label{subsec-4B}

In order to construct the cloaking potential $\check u_-(x)$ for the Bragg reflector potential $v_b(x)$, we use the model given by (\ref{model-r}) and (\ref{model-l}) to determine a concrete realization of the unidirectional invisible potentials $\check u_{1-}(x)$ and $\check u_{2-}(x)$. In view of (\ref{q103}), we set
    \begin{align}
    &\check u_{1-}(x)=v^r_{\cR_1}(x), && \check u_{2-}(x)=v^l_{\cR_2}(x)=v^r_{-\cR_2^*}(x)^*,
    \label{w11}
    \end{align}
where
    \begin{align*}
    &\cR_1:=-R_b^l/T_b= -2.180072 - 1.480919\,i\\
    &\cR_2:=(1-T_b)/R_b^l= -1.355199 - 0.030912\,i.
    \end{align*}
We also take the following values for the parameters $m,n,$ and $\alpha$ which enter the expression for $v^r_{\cR_1}(x)$ and $v^r_{-\cR_2^*}(x)^*$, respectively.
    \begin{align*}
    & m_1=-38094, && n_1=250 && \alpha_1=4.199808\times 10^{-4},\\
    & m_2=-19048, && n_2=250 && \alpha_2=1.158825\times 10^{-4}.
    \end{align*}
These correspond to a pair of 133~$\mu$m-thick parallel planar slabs placed on the left of the original Bragg reflector slab such that the distance between neighboring slabs is 1\,cm.

\begin{figure}
	\begin{center}
	\includegraphics[scale=.5]{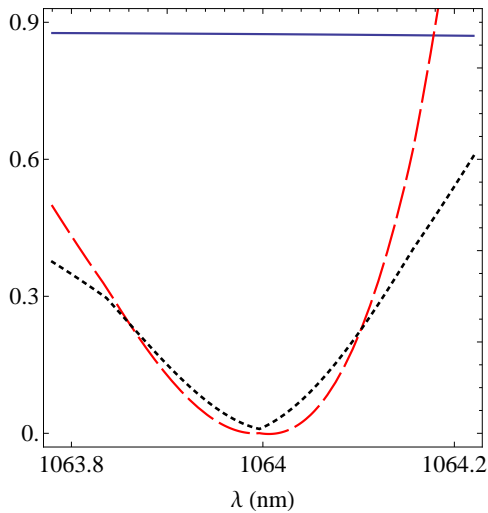}
	\caption{(Color online) Graph of $|R^l|^2$ for the Bragg reflector potential $v_b(x)$ (navy solid curve), and graphs of $|R^l|^2$ (red dashed curve) and $|T-1|$ (black dotted curve) for the potential $v_b(x)+\check u_-(x)$ of Sec.~\ref{subsec-4B}.}
	\label{fig3}
	\end{center}
	\end{figure}

Figure~\ref{fig3} shows the graphs of the left reflection coefficient for $v_b(x)$ before and after adding the cloaking potential,
    \be
    \check u_-(x)=\check u_{1-}(x)+\check u_{2-}(x)=v^r_{\cR_1}(x)+v^r_{-\cR_2^*}(x)^*,
    \label{w21}
    \ee
and the graph of $|T-1|$ for $v_b(x)+\check u_-(x)$. These graphs confirm the left-invisibility of $v_b(x)+\check u_-(x)$ within a very narrow spectral band centered at $\lambda=1064\,{\rm nm}$.

\subsection{Optimal invisibility cloaking potential $u_-(x)$}
\label{sub5c}

To construct the cloaking potential $u_-(x)$ for $v_b(x)$, we need to identify the potentials $u_{j- }(x)$ by fixing the values of the parameters $\cR, n, \alpha$, and $m$ in Eqs.~(\ref{model-r}) and (\ref{model-l}) appropriately. Let us label the values of these parameters for $u_{j-}(x)$ respectively by $\cR_{j},n_{j},\alpha_{j},$, and $m_{j}$, so that
    \begin{align*}
    & u_{1-}(x)=v_{\cR_1}^l(x)=v_{-\cR_1^*}^{r}(x)^*,\\
    & u_{2-}(x)=v_{\cR_2}^r(x),\\
    & u_{3-}(x)=v_{\cR_3}^l(x)=v_{-\cR_3^*}^{r}(x)^*.
    \end{align*}
Then according to (\ref{Mj-1}), (\ref{Mj-2}), (\ref{RL=b}), and (\ref{RRT=b}),
    \begin{align*}
    &\cR_{1}=T_b(T_b-1)/R_b^l= -0.333601 + 0.346363\,i, \\
    &\cR_{2}=-R_b^l/T_b= -2.180072 - 1.480919\,i, \\
    &\cR_{3}=(1-T_b)/R_b^l=  -1.355199 - 0.030912\,i.
    \end{align*}

If we take $n_{1}=n_2=n_3=250$, the potentials $u_{j-}(x)$ correspond to active slabs of thickness $133~\mu{\rm m}$, and we find
    \begin{align*}
    &\alpha_{1}=7.655383\times 10^{-5}, && m_1=-57142,\\
    &\alpha_{2}=4.199808\times 10^{-4}, && m_2=-38094,\\
    &\alpha_{3}=2.158825\times 10^{-4}, && m_3=-19048,
    \end{align*}
where we have made use of (\ref{e01}) and (\ref{d=}) and chosen $m_{j}$ in such a way that the distance between neighboring slabs associated with the potentials  $u_{j\pm}(x)$ and $v_b(x)$ is $1\,{\rm cm}$.

\begin{figure}[t]
	\begin{center}\vspace{.5cm}
	\includegraphics[scale=.5]{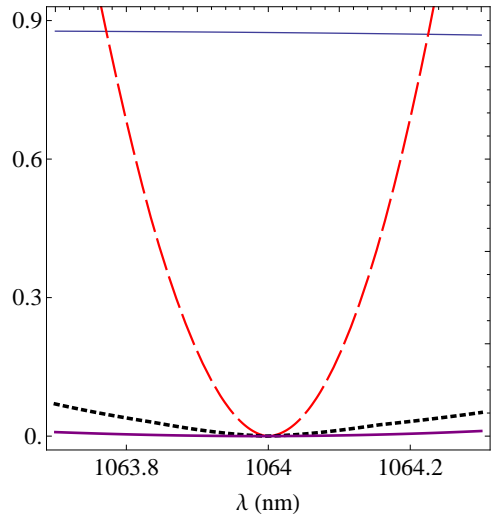}
	\caption{(Color online) Graph of $|R^l|^2$ for the Bragg reflector potential $v_b(x)$ of (\ref{bragg}) (thin navy solid curve), graphs of $|R^l|^2$ (red dashed curve) and $|T-1|$ (black dotted curve) for the cloaked potential $v_b(x)+u_-(x)$ of Sec.~\ref{sub5c}, and the graph of $|R^r-R_b^r|$ (thick purple curve), where $R^r$ and $R_b^r$ are respectively the right reflection amplitude of $v_b(x)+u_-(x)$ and $v_b(x)$.}
	\label{fig4}
	\end{center}
	\end{figure}

Figure~\ref{fig4} provides a graphical demonstration of the effect of adding the cloaking potential $u_-(x)$ to the Bragg reflector potential $v_b(x)$.
It confirms that indeed $v_b(x)+ u_-(x)$ is left-invisible and has the same right reflection amplitude as $v_b(x)$. Our numerical calculations show that, for $\lambda=1064\,{\rm nm}$, the reflection and transmission amplitudes of $v_b(x)+u_-(x)$ satisfy: $|R^l|^2<4.1\times 10^{-4}$, $|T-1|<5.0\times 10^{-5}$, and $|R^r-R^r_b|< 2.6 \times 10^{-5}$.

\section{Concluding Remarks}

The discovery that certain complex scattering potentials can display unidirectional invisibility is important, for they possess nonreciprocal scattering features which might find important applications in optical circuitry. This constituted the initial motivation for the study of these potentials \cite{lin,invisible-2}. Recently, it became clear that unidirectional invisible potentials played a central role in scattering theory, in the sense that they could be used to construct scattering potentials with any desired scattering properties at any prescribed wavenumber \cite{pra-2014b}.

In the present article we have employed this property of unidirectional invisible potentials to outline a method for the construction of various cloaking potentials which make a given finite-range real or complex potential unidirectionally reflectionless or invisible at an arbitrary wavenumber of our choice. In particular, we obtain optimal antireflection and invisibility cloaks which cannot be constructed using absorbing potentials. Our method relies on the use of up to three finite-range unidirectionally invisible potentials for which we give explicit close-form expressions.

In the application of our method for the Bragg reflector, the cloaking action is confined to a spectral band centered at the desired wavelength which can be as narrow as a small fraction of a nanometer. However, we should like to stress that the width of this band can be increased by a proper choice of the constituent unidirectionally invisible potentials. In this article, we have confined our attention to the class of unidirectionally invisible potentials (\ref{model-r}) and made essentially arbitrary choices for their parameters. Our method should, in principle, allow for broadband cloaking action, if we can supplement it with an optimization procedure to select the optimal choices for the constituent unidirectionally invisible potentials. This is a subject of a future investigation.

Finally, we wish to point out that the optimal antireflection and invisibility cloaks should offer useful applications in optical modulation and even display technologies, for they can be used as wavelength-sensitive one-way control devices.

\vspace{6pt}

\noindent{\em Acknowledgments:}  I would like to thank Aref Mostafazadeh for fruitful discussions. This work has been supported by  the Scientific and Technological Research Council of Turkey (T\"UB\.{I}TAK) in the framework of the project no: 112T951, and by the Turkish Academy of Sciences (T\"UBA).

\ed




\begin{thebibliography}{99}
\bibitem{f1} The support of a function $v(x)$ of a real variable $x$ is the smallest closed interval outside which $v(x)$ vanishes.
\end{thebibliography}
